\documentstyle[aps,twocolumn,eqsecnum]{revtex}
\begin{document}
\draft
\title{Thermal Green Functions in Coordinate Space
 for Massless Particles of any Spin}
\author{H. Arthur Weldon}
\address{Department of Physics, West Virginia University, Morgantown WV
26506-6315}
\date{February 2, 2000}

\maketitle
\begin{abstract}
The  thermal Wightman functions  for free, massless particles of
spin 0, 1/2, 1,  3/2, and 2 are computed directly in coordinate space by
solving the appropriate differential equation and imposing the
Kubo-Martin-Schwinger condition. The solutions are valid for real,
imaginary, or complex time. The
Wightman functions for spin 1 gauge bosons and for spin 2 gravitons are
directly related to the fundamental functions for spin 0. The Wightman
functions for spin 3/2 gravitinos is directly related to  that for spin 1/2
fermions.   Calculations for spin 1, 3/2, and 2 are done in covariant
gauges.   In the deep space-like region  the
Wightman functions for bosons fall like $T/r$ whereas those for the fermions
fall exponentially. In the deep time-like region all the Wightman functions
fall
exponentially.
\end{abstract}

\def\sqr#1#2{{\vcenter{\vbox{\hrule height.#2pt
	\hbox{\vrule width.#2pt  height#1pt   \kern#1pt \vrule  width.#2pt}
	\hrule height.#2pt}}}}

\section{Introduction}

In both zero-temperature and finite-temperature field theory it is
customary to
perform  calculations in momentum space.
There are exceptions to this pattern however. Lattice gauge theory
computations are done in coordinate space. The short-distance
operator product expansion is formulated in coordinate space.  It
is even possible to carry out ultraviolet regularization and
renormalization directly in coordinate space
\cite{uv}.

The purpose of this paper is to deduce the free Wightman functions
at finite temperature
for  various massless particles directly in coordinate space.
Thermal averages are performed with respect to the equilibrium density
operator
\begin{equation}
\varrho=e^{-\beta H}/{\rm Tr}[e^{-\beta H}],
\end{equation}
where $H$ is the appropriate Hamiltonian and $\beta=1/T$.
Let $\phi^{A}(x)$  denote a quantum field of any spin. For spin-zero fields
the index $A$ only distinguishes species. For spin 1/2 fields $A$ denotes a
spinor index; for spin 1, a vector index; for spin 3/2 a spinor and vector
index; for spin 2, a pair of  vector indices.
The  Wightman
 functions at finite temperature are
\begin{eqnarray}
G_{>}^{AB}(x)=&&-i{\rm Tr}\big(\varrho \phi^{A}(x)\phi^{B}(0)\big)
\nonumber\\
G_{<}^{AB}(x)=&&-i{\rm Tr}\big(\varrho \phi^{B}(0)\phi^{A}(x)\big)
(-1)^{2J}.
\label{Wight}\end{eqnarray}
Knowing the thermal Wightman function allows  direct construction of
 the various thermal propagators in real or imaginary
time \cite{b1,b2,b3}.  In particular the time-ordered Green function is
\begin{equation}
G^{AB}(x)=\theta(t)G_{>}^{AB}(x)
+\theta(-t)G_{<}^{AB}(x).
\end{equation}

The canonical method for obtaining $G_{>}(x)$ for
free fields would require the following steps: (1) Solve the free field
equations and express the  field operator in terms of plane-wave
solutions weighted by creation and annihilation operators. (2) Impose
the equal-time canonical commutation relations. (3)
Express the Hamiltonian in terms of creation and annihilation
operators. (4) Compute the  trace in Fock space over the density
operator. (5) Perform the integral over momentum states
so as to obtain $G_{>}(x)$  in coordinate space.
For higher spins this becomes tedious.

The first four steps can  be avoided if the $T=0$
propagator is already known, because the  spectral function
$\rho^{AB}(K)$ gives directly the Wightman function in momentum space
\cite{b1,b2,b3}:
\begin{equation}
G^{AB}_{>}(K)={-i\rho^{AB}(K)\over 1\pm e^{-\beta k_{0}}}.\end{equation}
However it is still necessary to  Fourier transform from momentum space to
coordinate space and obtain $G_{>}(x)$. This was the procedure
followed in
\cite{HAW}, which computed the coordinate-space Wightman functions for
massless vector  bosons in various gauges (Feynman, general covariant,
and Coulomb). However performing the Fourier
transforms so as to ensure the correct analyticity properties in
complex time is difficult.  For gauge bosons the complete answer
satisfying all the analyticity properties was only obtained in the
Feynman gauge.
It would be laborious to pursue the Fourier transform method for spin
1/2 fermions, for gravitinos, or for gravitons.

It turns out to be simpler to deduce the thermal
Wightman functions  for free fields by working directly in
coordinate space.  Since $G_{>}(x)$ solves the free
field equation all that is necessary is that the solution have the correct
zero-temperature limit and  satisfy the  Kubo-Martin-Schwinger
periodicity relation
\cite{b1,b2,b3,KMS} under $t\to t-i\beta$.
It has  not been generally recognized  that the
KMS relation is not only a necessary condition that thermal Wightman
functions must satisfy but is also sufficient condition to determine them
directly.

The paper is organized according to spin. Sec II deals with bosons.  The
thermal Wightman functions for spinless bosons is given in Eq.
(\ref{2aD>D<}), for gauge bosons in Eq. (\ref{2bD}), and for  gravitons in
Eq. (\ref{2cD}) and (\ref{2cgrav}). Sec III deals with fermions. The thermal
Wightman functions for spin 1/2 fermions is given by Eq.
(\ref{3aS>S<}) and (\ref{3as>s<}); for spin 3/2 gravitinos in Eq.
(\ref{3bL}) and (\ref{3bS>S<}). In the covariant gauges considered here, the
higher spin functions are all expressible in terms of the basic spin 0 and
1/2 functions. Sec III discusses the asymptotic behavior.

Throughout the paper the fundamental
functions depend only radial distance $r$ and time $t$ and it is convenient
to use variables $u$ and $v$ defined as follows:
\begin{equation}
u=r+t\hskip1cm v=r-t.\label{uv}\end{equation}
To avoid confusion with the metric $g_{\mu\nu}$ in curved
space, the Minkowski metric will everywhere be denoted by
$\eta_{\mu\nu}$.

\section{Bosons of spin 0,1,2}

\subsection{Spinless Bosons}

For a spinless boson field $\phi(x)$
the basic thermal  Wightman functions are
\begin{eqnarray}
D_{>}(x)=&&-i{\rm Tr}\big(\varrho\, \phi(x)\phi(0)\big)\nonumber\\
D_{<}(x)=&&-i{\rm Tr}\big(\varrho\, \phi(0)\phi(x)\big).\label{2aWightman}
\end{eqnarray}
The emphasis of the subsequent development will be to avoid expressing
the free field operator as a sum of plane waves weighted by
creation and annihilation operators.
It will be more direct to solve the  Klein-Gordon equation
$\sqr66\,D_{>}(x)=\sqr66\,D_{<}(x)=0$ subject to various conditions.
One such constraint is the
 normalization condition provided by imposing
the canonical value of the equal-time commutator
$[\dot{\phi},\phi]$. This requires
\begin{equation}
{\partial\over \partial t}\Big(D_{>}(x)-D_{<}(x)\Big)\Big|_{t=0}
=-\delta^{3}(\vec{r}).\label{2atime}\end{equation}

The zero-temperature solution to the homogeneous differential
equation satisfying the above initial condition is
\begin{eqnarray}
D_{>}(x)\big|_{T=0}&&={i\over 4\pi^{2}}{1\over (t-i\epsilon)^{2}-r^{2}}
\nonumber\\
D_{<}(x)\big|_{T=0}&&={i\over 4\pi^{2}}{1\over (t+i\epsilon)^{2}-r^{2}}
.\nonumber\end{eqnarray}
At zero temperature $D_{>}(x)$ is analytic throughout the lower-half of the
complex
$t$ plane and $D_{<}(x)$ is analytic throughout the upper-half of the
complex $t$
 plane.  For later purposes it will be convenient to write these in terms of
the variables $u$ and $v$ of Eq. (\ref{uv}):
\begin{eqnarray}
D_{>}(x)\big|_{T=0}&&={-i\over 8\pi^{2}r}\bigg[{1\over u-i\epsilon}
+{1\over v+i\epsilon}\bigg]
\nonumber\\
D_{<}(x)\big|_{T=0}&&={-i\over 8\pi^{2}r}\bigg[{1\over u+i\epsilon}
+{1\over v-i\epsilon}\bigg]
\end{eqnarray}

The finite-temperature  Wightman functions  have some standard
properties \cite{b1,b2,b3} that are easily demonstrated by inserting a
complete set of energy eigenstates between the operators in
(\ref{2aWightman}). First,
$D_{>}(x)$ is an analytic function of complex time in the  strip
  $-\beta\le{\rm
Im}\,t\le 0$, which will be referred to as the lower
strip. Similarly $D_{>}(x)$ is analytic in complex time in the  strip
$0\le {\rm Im}\,t\le \beta$,  referred to as the upper strip.
The values of each Wightman functions on the boundaries of its region of
analyticity are related by the Kubo-Martin-Schwinger (KMS)
conditions \cite{b1,b2,b3,KMS}:
\begin{eqnarray}
D_{>}(t-i\beta,r)=&&D_{<}(t,r)\nonumber\\
D_{<}(t+i\beta,r)=&&D_{>}(t,r)\label{2aKMS}
\end{eqnarray}

The special property of free fields that will be essential is that their
commutator
$\big[\phi(x),\phi(0)\big]$ is a c-number at all $x$.
The difference of the two Wightman functions is the thermal average of this
c-number and  therefore the difference is independent of  temperature:
\begin{equation}
D_{>}(x)-D_{<}(x)=D_{>}(x)\big|_{T=0}-D_{<}(x)\big|_{T=0}
.\end{equation}
This implies that the Wightman functions have the structure
\begin{eqnarray}
D_{>}(x)=D_{>}(x)\big|_{T=0}+E(x)\nonumber\\
D_{<}(x)=D_{<}(x)\big|_{T=0}+E(x).\nonumber
\end{eqnarray}
Since $D_{>}(x)$ is analytic in the lower strip and $D_{<}(x)$ analytic in
the upper strip, the function $E(x)$ is actually analytic in
the double-width strip $-\beta\le{\rm Im}\,t\le\beta$.

Because  the fields satisfy the massless Klein-Gordon
equation,  then $E(x)$ does too:
$\sqr66\,E(x)=0$.
The most general spherically symmetric solution has the form
\begin{displaymath}
E(x)={1\over r}\Big(F_{1}(u)+F_{2}(v)\Big).
\end{displaymath}
From the definitions in Eq. (\ref{2aWightman}), time translation invariance
implies that
$D_{>}(t,r)=D_{<}(-t,r)$. Consequently
$E(x)$ must be an even function of time. This makes the two
functions are the same: $F_{1}=F_{2}$.
Therefore the Wightman functions have the form
\begin{eqnarray}
D_{>}(x)={-i\over 8\pi^{2}r}\bigg[&&{1\over u-i\epsilon}
+F(u)+{1\over v+i\epsilon}+F(v)\bigg]\nonumber\\
D_{<}(x)={-i\over 8\pi^{2}r}\bigg[&&{1\over u+i\epsilon}
+F(u)+{1\over v-i\epsilon}+F(v)\bigg].\label{D>D<}
\end{eqnarray}
The KMS condition Eq. (\ref{2aKMS}) requires that $F$ satisfy
\begin{equation}
{1\over u\!+\!i\epsilon}+F(u)={1\over u\!-\!i\beta\!-\!i\epsilon}
+F(u\!-\!i\beta).
\end{equation}
This fixes $F(u)$ to have an infinite number of simple poles:
\begin{equation}
F(u)=\sum_{n=1}^{\infty}\bigg[{1\over u\!-\!i(n\beta\!+\!\epsilon)}
+{1\over u\!+\!i(n\beta\!+\!\epsilon)}\bigg].
\label{2aSeries}
\end{equation}
This is the complete answer and has all the correct properties.
  Since
${\rm Im}\,u={\rm Im}\,t$, there are no singularites in the closed
strip
$-\beta\le{\rm Im}\,t\le\beta$ as expected.  The nearest poles are
outside this strip at $u=\pm i(\beta+\epsilon)$. It is relatively easy
to sum this series and obtain a more useful expression.

\underbar{Complex time in the open strip:}
 Often one is interested in complex time in the
open strip
\begin{equation}
-\beta<{\rm Im}\,t<\beta,\label{2aStrip}\end{equation}
which of course includes real time.
For the open strip, one can set
$\epsilon\to\! 0$. Then the series sums to
\begin{displaymath}
F(u)=-{1\over u}+\pi T\coth(\pi Tu).
\end{displaymath}
This has no pole at $u=0$ but does have poles at the border $u=\pm i\beta$
because of the limit taken. Substituting this into Eq. (\ref{D>D<})
gives
\begin{mathletters}\begin{eqnarray}
D_{>}(t,r)=&&{1\over 8\pi r}\big(\delta(u)-\delta(v)\big)\nonumber\\
-&&{iT\over 8\pi r}\big(\coth(\pi Tu)+\coth(\pi T v)\big)\\
D_{<}(t,r)=&&{1\over 8\pi r}\big(-\delta(u)+\delta(v)\big)\nonumber\\
-&&{iT\over 8\pi r}\big(\coth(\pi Tu)+\coth(\pi T v)\big)\end{eqnarray}
\label{2aD>D<}\end{mathletters}
Note that it is only possible for $t$
and
$t-i\beta$ to  both be in the open strip defined in Eq. (\ref{2aStrip})
if both are complex. But if both are complex then the Dirac delta functions
have no support. Consequently the KMS condition is satisfied trivially here.
It will be satisfied in a nontrivial fashion below in Eq. (\ref{d>d<}).

\underbar{Imaginary time:} The imaginary time formalism uses $t=-i\tau$ where
$\tau$ is real. Then $u=r-i\tau$ and $v=r+i\tau$. The two Wightman functions
are equal, $D_{>}(-i\tau,r)=D_{<}(-i\tau,r)$ and given by
\begin{equation}
D(-i\tau,r)={-iT\over 4\pi r}{\sinh(2\pi Tr)\over
\cosh(2\pi Tr)-\cos(2\pi T\tau)}.
\end{equation}
Naturally this is periodic under $\tau\to\tau\pm \beta$.

\underbar{Arbitrary complex time:} To treat all the poles
of $F(u)$ in Eq. (\ref{2aSeries})
 correctly one cannot use the limit $\epsilon\to 0$.
For non-zero $\epsilon$ the series can be summed in terms of the standard
function
$\psi(z)=d\ln[\Gamma(z)]/dz$ to obtain
\begin{displaymath}
F(u)=iT\psi[1\!+\!T(\epsilon\!-\!iu)]
-iT\psi[1\!+\!T(\epsilon\!+\!iu)].
\end{displaymath}
This is analytic in the closed strip $-\beta\le
{\rm Im}\,t\le\beta$ as expected.

\underbar{Concise notation:}
For later purposes it is convenient to express $F(u)$  as a
derivative. Therefore define a new function
\begin{equation}
f(u)= \Gamma[1\!+\!T(\epsilon\!-\!iu)]\Gamma[1\!+\!T(\epsilon\!+\!iu)].
\end{equation}
Then $F(u)=-\partial\ln[f(u)]/\partial r$
so that
\begin{displaymath}
{1\over r}F(u)=-{1\over r}{\partial\over\partial r}
\ln[f(u)]
={1\over 2}\,\sqr66\,\ln[f(u)].
\end{displaymath}
Comparison with Eq. (\ref{D>D<}) shows that the  thermal
Wightman functions can be concisely expressed as
\begin{mathletters}\begin{eqnarray}
D_{>}(x)=&&\sqr66\;d_{>}(x)\\
D_{<}(x)=&&\sqr66\;d_{<}(x).
\end{eqnarray}\end{mathletters}
The lower case functions $d(x)$ are given by
\begin{mathletters}\begin{eqnarray}
d_{>}(x)=&&-{i\over 16\pi^{2}}\ln\bigg[{f(u)f(v)\over
(u-i\epsilon)(v+i\epsilon)}
\bigg]\\
d_{<}(x)=&&-{i\over 16\pi^{2}}\ln\bigg[{f(u)f(v)\over
(u+i\epsilon)(v-i\epsilon)}
\bigg]\end{eqnarray}\label{d>d<}\end{mathletters}
Note that the difference,  $d_{>}(x)-d_{<}(x)$, is
independent of temperature.

\underbar{Special case:}
For many purposes, such as real time, it is
adequate to use the open strip given in Eq. (\ref{2aStrip}), which results
from
$\epsilon\to 0$. In this limit
\begin{equation}
\lim_{\epsilon\to 0}f(u)={\pi T u\over\sinh(\pi T u)}.
\end{equation}
This makes Eq. (\ref{d>d<}) rather simple. The $\pm i\epsilon$
remaining in Eq. (\ref{d>d<}) produce the correct light-cone
singularities
$\delta(u)$ and $\delta(v)$ displayed earlier. In two important cases
these Dirac delta functions have no support: either $t$ is complex or
$t$ is real but not on the light-cone. In either case one can omit the
$\pm i\epsilon$ in Eq. (\ref{d>d<}) in which case the two functions
are equal: $d_{>}(x)=d_{<}(x)\equiv d(x)$ where
\begin{equation}
d(x)={i\over
16\pi^{2}}\ln\big[\sinh(\pi Tu)\sinh(\pi Tv)\big].
\end{equation}
In this regime the two thermal Wightman functions are equal:
$D_{>}(x)=D_{<}(x)=\sqr66\,d(x)$.

\subsection{Spin 1 Gauge Bosons}

The same methods can be used to obtain the thermal Wightman functions
for massless gauge bosons directly in coordinate space.
In a general covariant gauge the Lagrangian density is
\begin{displaymath}
{\cal L}=-{1\over 4}F_{\mu\nu}F^{\mu\nu}
-{1\over 2\xi}(\partial_{\mu}A^{\mu})^{2},
\end{displaymath}
where $F_{\mu\nu}=\partial_{\mu}A_{\nu}-\partial_{\nu}A_{\mu}$ and $\xi$ is
an arbitrary gauge parameter. The thermal Wightman functions
are
\begin{eqnarray}
D^{\mu\nu}_{>}(x)&&=-i{\rm Tr}\big(\varrho\,
A^{\mu}(x)A^{\nu}(0)\big)\nonumber\\ D^{\mu\nu}_{<}(x)&&=-i{\rm
Tr}\big(\varrho\, A^{\nu}(0)A^{\mu}(x)\big).
\nonumber\end{eqnarray}
One can deduce these from first principles using the same arguments as
employed  in Sec IIA.
The subsequent discussion will show that
\begin{eqnarray}
D_{>}^{\mu\nu}(x)=&&
\big(-\eta^{\mu\nu}\sqr66+(1\!-\!\xi)\,
\partial^{\mu}\partial^{\nu}\big)d_{>}(x)\nonumber\\
D_{<}^{\mu\nu}(x)=&&\big(-\eta^{\mu\nu}\sqr66+(1\!-\!\xi)\,
\partial^{\mu}\partial^{\nu}\big)d_{<}(x),
\label{2bD}\end{eqnarray}
where $d_{>}(x)$ and $d_{<}(x)$ are the functions
 functions already given in Eq.  (\ref{d>d<}).

(1) The first check of these  Wightman functions is that
they have the correct analyticity in complex time and satisfy the correct
KMS condition
\begin{displaymath}
D_{>}^{\mu\nu}(t-i\beta,\vec{r})=D^{\mu\nu}_{<}(t,\vec{r}).
\end{displaymath}

(2) The next check is that the Wightman functions must satisfy the
correct  homogeneous differential equation. The variation of the
Lagrangian gives
\begin{equation}
{\partial{\cal L}\over \partial(\partial^{\rho}A^{\kappa})}
=P_{\rho\kappa\mu}A^{\mu},
\label{2bP1}\end{equation}
where the tensor  $P$ is linear in the first derivative:
\begin{equation}
P_{\rho\kappa\mu}=\eta_{\rho\mu}\partial_{\kappa}-\eta_{\kappa\mu}
\partial_{\rho}
-{1\over\xi}\eta_{\rho\kappa}\partial_{\mu}.\label{2bP2}\end{equation}
The equation of motion for the field is
\begin{displaymath}
0=\partial^{\rho}P_{\rho\kappa\mu}A^{\mu}=
\big[-\eta_{\kappa\mu}\sqr66\,+(1\!-\!{1\over\xi})\partial_{\kappa}
\partial_{\mu}\big]A^{\mu}.
\end{displaymath}
Applying this differential operator to the Wightman function
in Eq. (\ref{2bD}) gives
\begin{displaymath}
\Big(-\eta_{\lambda\mu}\sqr66+(1\!-\!{1\over\xi})\partial_{\lambda}
\partial_{\mu}\Big)
D^{\mu\nu}_{>}(x)=-\delta_{\lambda}^{\;\nu}\,
\sqr66\,\sqr66\,d_{>}(x).
\end{displaymath}
From Sec IIA, $\sqr66\,\sqr66\,d_{>}(x)=\sqr66\,D_{>}(x)=0$ and so the
equation of motion is satisfied.

(3) The third check is that the field operators in the Wightman functions
satisfy the correct canonical commutation relations.
The canonical momentum conjugate to $A^{\kappa}$ is
\begin{displaymath}
\Pi_{\kappa}=P_{0\kappa\mu}A^{\mu}.
\end{displaymath}
The equal-time canonical commutation relations are
\begin{displaymath}
-i\big[\Pi_{\kappa}(x),A^{\nu}(0)\big]_{t=0}
=-\delta_{\kappa}^{\;\nu}\;\delta^{3}(\vec{r}).
\end{displaymath}
This requires that the  Wightman functions to satisfy
\begin{equation}
\Big[P_{0\kappa\mu}(D_{>}^{\mu\nu}(x)-D_{<}^{\mu\nu}(x))\Big]_{t=0}=
-\delta_{\kappa}^{\;\nu}\,\delta^{3}(\vec{r}).\end{equation}
To check this, use Eq. (\ref{2bD}) and (\ref{2bP2}) to obtain
\begin{displaymath}
P_{\rho\kappa\mu}D_{>}^{\mu\nu}(x)=(\delta_{\kappa}^{\nu}\partial_{\rho}
-\delta_{\rho}^{\nu}\partial_{\kappa}
+\eta_{\rho\kappa}\partial^{\nu})
D_{>}(x),\end{displaymath}
where $D_{>}(x)$ is the thermal Wightman function for scalars.
For the appropriate difference of Wightman functions one needs
\begin{eqnarray}
P_{0\beta\mu}\big(D_{>}^{\mu\nu}(x)&&-D_{<}^{\mu\nu}(x)\big)\nonumber\\
=&&(\delta_{\kappa}^{\;\nu}\partial_{0}
-\delta_{0}^{\;\nu}\partial_{\kappa}
+\eta_{0\kappa}\partial^{\nu})(D_{>}(x)-D_{<}(x)).
\nonumber\end{eqnarray}
The right hand side contains both time derivatives and space derivatives.
At $t=0$ the spatial derivatives vanish because $D_{>}(0,\vec{r})
=D_{<}(0,\vec{r})$. The above result simplifies to
\begin{displaymath}
\big[P_{0\beta\mu}\big(D_{>}^{\mu\nu}(x)-D_{<}^{\mu\nu}(x)\big)
\big]_{t=0}\!\!=\delta_{\kappa}^{\,\nu}\big[
(\dot{D}_{>}(x)-\dot{D}_{<}(x)\big]_{t=0}.
\end{displaymath}
Eq. (\ref{2atime}) determines that the value of the right hand
side is
$-\delta_{\kappa}^{\;\nu}
\delta^{3}(\vec{r})$, as required. This completes the proof that
Eq. (\ref{2bD}) is correct.

\subsection{Spin 2 Gravitons}

Standard quantum gravity is based on the Einstein-Hilbert
Lagrangian  with a gauge-fixing term
\begin{displaymath}
{\cal L}={2\over \kappa^{2}}\sqrt{\!-\!g}\;R+{\cal L}_{\rm g.f.}
\end{displaymath}
in which $R$ is the scalar curvature, $g={\rm det}(g_{\mu\nu})$, and
$\kappa^{2}=32\pi G$ with $G$ Newton's constant. A conventional gauge
fixing term is \cite{G1,G2}
\begin{displaymath}
{\cal L}_{\rm g.f.}={1\over k^{2}}\eta_{\mu\nu}(\partial_{\alpha}
\sqrt{\!-\!g}g^{\alpha\mu})(\partial_{\beta}\sqrt{\!-\!g}g^{\beta\nu}).
\end{displaymath}
This corresponds to the Feynman gauge in Yang-Mills theories. More
general covariant gauges will be discussed later.
With $\eta^{\mu\nu}$  the  Minkowski metric,  the
graviton field $h^{\mu\nu}$ contains all quantum fluctuations:
\begin{equation}
\sqrt{\!-\!g}g^{\mu\nu}=\eta^{\mu\nu}+\kappa h^{\mu\nu}.
\end{equation}
Keeping only terms that are quadratic in $h$ produces the free Lagrangian
density
\begin{eqnarray}
{\cal L}_{0}={1\over 2}&&(\partial_{\rho}h_{\alpha\beta})
(\partial^{\rho}h^{\alpha\beta})
-{1\over 4}(\partial_{\rho}h_{\alpha}^{\;\;\alpha})
(\partial^{\rho}h_{\beta}^{\;\;\beta})\nonumber\\
-&&(\partial_{\rho}h_{\alpha\beta})
(\partial^{\alpha}h^{\rho\beta})+
(\partial_{\rho}h^{\rho\alpha})(\partial^{\beta}h_{\beta\alpha}).\nonumber
\end{eqnarray}
The thermal Wightman functions to be computed are
\begin{eqnarray}
D_{>}^{\mu\nu\alpha\beta}(x)=&&-i{\rm Tr}\,\big(\varrho\,h^{\mu\nu}(x)
h^{\alpha\beta}(0)\big)\nonumber\\
D_{<}^{\mu\nu\alpha\beta}(x)=&&-i{\rm Tr}\,\big(\varrho\,
h^{\alpha\beta}(0)h^{\mu\nu}(x)\big).\nonumber
\end{eqnarray}
Subsequent argument will show that these are given by
\begin{eqnarray}
D_{>}^{\mu\nu\alpha\beta}(x)\!=&&\big(\!-\eta^{\mu\nu}\eta^{\alpha\beta}
\!+\!\eta^{\mu\alpha}\eta^{\nu\beta}
\!+\!\eta^{\mu\beta}\eta^{\nu\beta}\big)
D_{>}(x)\nonumber\\
D_{<}^{\mu\nu\alpha\beta}(x)\!=&&\big(\!-\eta^{\mu\nu}\eta^{\alpha\beta}
\!+\!\eta^{\mu\alpha}\eta^{\nu\beta}\!+\!\eta^{\mu\beta}\eta^{\nu\beta}
\big)D_{<}(x).\!\label{2cD}
\end{eqnarray}
That the results are expressed in terms of the scalar Wightman
functions $D_{>}(x)$  rather than the potential functions $d_{>}(x)$
is a peculiarity of this Feynman-like gauge. In more general covariant
gauges the Wightman functions for gravitons depends on the potentials
$d_{>}(x)$.

(1) Because of the properties of the scalar Wightman functions, the
graviton Wightman functions in Eq. (\ref{2cD}) satisfy
\begin{displaymath}
D^{\mu\nu\alpha\beta}_{>}(t-i\beta,\vec{r})
=D^{\mu\nu\alpha\beta}_{>}(t,\vec{r}),
\end{displaymath}
and are analytic in the appropriate regions.

(2) To obtain the equation for the graviton field one needs the
partial derivatives
\begin{eqnarray}
{\partial{\cal L}\over\partial(\partial^{\rho}h^{\kappa\lambda})}
=P_{\rho\kappa\lambda\mu\nu}h^{\mu\nu},
\end{eqnarray}
where the tensor $P$ is linear in the derivative operator
\begin{eqnarray}
P_{\rho\kappa\lambda\mu\nu}=&&(\Lambda_{\kappa\lambda\mu\nu}-{1\over
2}\eta_{\kappa\lambda}\eta_{\mu\nu})\partial_{\rho}\nonumber\\
-&&\Lambda_{\rho\lambda\mu\nu}\,\partial_{\kappa}
-\Lambda_{\kappa\rho\mu\nu}\,\partial_{\lambda}\label{2cP1}\\
+&&\Lambda_{\kappa\lambda\rho\nu}\,\partial_{\mu}
+\Lambda_{\kappa\lambda\mu\rho}\,\partial_{\nu}\nonumber
\end{eqnarray}
and $\Lambda$ is given by
\begin{equation}
\Lambda_{\kappa\lambda\mu\nu}={1\over
2}(\eta_{\kappa\mu}\eta_{\lambda\nu}
+\eta_{\kappa\nu}\eta_{\lambda\mu}).
\end{equation}
The  differential equation for the graviton field operator
is $\partial^{\rho}P_{\rho\kappa\lambda\mu\nu}h^{\mu\nu}(x)=0$, or more
concisely
\begin{equation}
(-\eta_{\kappa\lambda}\eta_{\mu\nu}
+\eta_{\kappa\mu}\eta_{\lambda\nu}+\eta_{\kappa\nu}\eta_{\lambda\mu})\,
\sqr66\,h^{\mu\nu}(x)=0.
\end{equation}
The graviton Wightman functions Eq. (\ref{2cD}) automatically satisfy
$\sqr66\,D^{\mu\nu\alpha\beta}_{>}(x)=0$ since
the scalar functions satisfy $\sqr66\,D_{>}(x)=0$.

(3) The third check of Eq.
(\ref{2cD}) is that  the graviton operators  satisfy the
correct canonical commutation relations. The canonical momentum conjugate
to $h^{\kappa\lambda}$ is
\begin{displaymath}
\Pi_{\kappa\lambda}=P_{0\kappa\lambda\mu\nu}h^{\mu\nu}.
\end{displaymath}
The canonical equal-time commutation relations are
\begin{displaymath}
-i\big[\Pi_{\kappa\lambda}(x),h^{\alpha\beta}(0)\big]_{t=0}
=-(\delta_{\kappa}^{\;\alpha}\delta_{\lambda}^{\;\beta}
+\delta_{\kappa}^{\;\beta}\delta_{\lambda}^{\;\alpha})
\delta^{3}(\vec{r}).
\end{displaymath}
Therefore the Wightman functions should  satisfy
\begin{eqnarray}
\big[P_{0\kappa\lambda\mu\nu}\big(D_{>}^{\mu\nu\alpha\beta}(x)
&&-D_{<}^{\mu\nu\alpha\beta}(x)\big)\big]_{t=0}\nonumber\\
=-&&(\delta_{\kappa}^{\;\alpha}\delta_{\lambda}^{\;\beta}
+\delta_{\kappa}^{\;\beta}\delta_{\lambda}^{\;\alpha})
\delta^{3}(\vec{r}).\label{2cMom}
\end{eqnarray}
To check that the Wightman function  Eq. (\ref{2cD}) satisfies this,
first apply the differential operator in Eq. (\ref{2cP1}):
\begin{eqnarray}
P_{\rho\kappa\lambda\mu\nu}\big(D_{>}^{\mu\nu\alpha\beta}&&(x)
-D_{>}^{\mu\nu\alpha\beta}(x)\big)\nonumber\\
=&&(\delta_{\kappa}^{\;\alpha}\delta_{\lambda}^{\;\beta}
+\delta_{\kappa}^{\;\beta}\delta_{\lambda}^{\;\alpha})
\partial_{\rho}\big(D_{>}(x)\!-\!D_{<}(x)\big)\nonumber\\
-&&(\delta_{\rho}^{\;\alpha}\delta_{\lambda}^{\;\beta}
+\delta_{\rho}^{\;\beta}\delta_{\lambda}^{\;\alpha})
\partial_{\kappa}\big(D_{>}(x)\!-\!D_{<}(x)\big)\nonumber\\
-&&(\delta_{\kappa}^{\;\alpha}\delta_{\rho}^{\;\beta}
+\delta_{\kappa}^{\;\beta}\delta_{\rho}^{\;\alpha})
\partial_{\lambda}\big(D_{>}(x)\!-\!D_{<}(x)\big)\nonumber\\
+&&(\eta_{\rho\kappa}\,\delta_{\lambda}^{\;\beta}
+\eta_{\rho\lambda}\delta_{\kappa}^{\;\beta})
\partial^{\alpha}\big(D_{>}(x)\!-\!D_{<}(x)\big)\nonumber\\
+&&(\eta_{\rho\kappa}\,\delta_{\lambda}^{\;\alpha}
+\eta_{\rho\lambda}\,\delta_{\kappa}^{\;\alpha})
\partial^{\beta}\big(D_{>}(x)\!-\!D_{<}(x)\big).\nonumber
\end{eqnarray}
Now set the index $\rho=0$ and
the time $t=0$. The first line obviously coincides with Eq. (\ref{2cMom})
and it is straightforward to check that
 the remaining four lines will always vanish. For example, if
all of the free indices $\kappa,\lambda,
\mu,\nu$ are spatial, the last four lines vanish. This, because the  spatial
derivatives on the right hand side are zero since
$D_{>}(0,\vec{r})=D_{<}(0,\vec{r})$. If three of the indices
$\kappa,\lambda,\mu,\nu$ are spatial and one is 0, there will be one
non-vanishing time derivatives among the last four lines but it will be
multiplied by a tensor that vanishes.  For the remaining cases it is easy to
enumerate the possible values of the free indices
$\kappa,\lambda,\mu,\nu$ and verify that the canonical commutation
relation Eq. (\ref{2cMom}) is fully satisfied.

\underbar{General Covariant Gauge}: The results for the Wightman function
given in Eq. (\ref{2cD}) are for a particular gauge analogous to the
Feynman gauge. A more general gauge-fixing term is
\begin{displaymath}
{\cal L}_{\rm g.f.}={1\over\xi k^{2}}\eta_{\mu\nu}(\partial_{\alpha}
\sqrt{\!-\!g}g^{\alpha\mu})(\partial_{\beta}\sqrt{\!-\!g}g^{\beta\nu}).
\end{displaymath}
where $\xi$ is arbitrary. The differential equation for the graviton
field $h^{\mu\nu}(x)$ is then
\begin{eqnarray}
&&(-\eta_{\kappa\lambda}\eta_{\mu\nu}
+\eta_{\kappa\mu}\eta_{\lambda\nu}+\eta_{\kappa\nu}\eta_{\lambda\mu})\,
\sqr66\,h^{\mu\nu}\nonumber\\
&&=(1\!-\!{1\over \xi})(\eta_{\kappa\mu}\partial_{\lambda}\partial_{\nu}
\!+\!\eta_{\kappa\nu}\partial_{\lambda}\partial_{\mu}
\!+\!\eta_{\lambda\mu}\partial_{\kappa}\partial_{\nu}
\!+\!\eta_{\lambda\nu}\partial_{\kappa}\partial_{\mu})
h^{\mu\nu}.\nonumber\end{eqnarray}
Following the same procedures as above one can  show that
the thermal Wightman function is
\begin{eqnarray}
D_{>}^{\mu\nu\alpha\beta}(x)&&=
\big(\!-{1\over\xi}\eta^{\mu\nu}\eta^{\alpha\beta}
\!+\!\eta^{\mu\alpha}\eta^{\nu\beta}
\!+\!\eta^{\mu\beta}\eta^{\nu\beta}\big)
\sqr66\,d_{>}(x)\nonumber\\
+(1\!-\!{1\over\xi})&&\Big\{\!-2
(\eta^{\mu\nu}\partial^{\alpha}\partial^{\beta}
+\eta^{\alpha\beta}\partial^{\mu}\partial^{\nu})d_{>}(x)\label{2cgrav}\\
+(\eta^{\mu\alpha}&&\partial^{\nu}\partial^{\beta}
\!+\!\eta^{\mu\beta}\partial^{\nu}\partial^{\alpha}
\!+\!\eta^{\nu\alpha}\partial^{\mu}\partial^{\beta}
\!+\!\eta^{\nu\beta}\partial^{\mu}\partial^{\alpha})d_{>}(x)
\Big\}.\nonumber
\end{eqnarray}
Note that the general Wightman function depends on all partial
derivatives of $d_{>}(x)$, whereas Eq. (\ref{2cD}), in the Feynman-like
gauge,  depends only on
$D_{>}(x)=\sqr66\,d_{>}(x)$.

\section{Fermions}

The methods employed in the previous section carry over to fermions except
that
the KMS condition contains an
additional  minus sign. This change results in a completely different
asymptotic
behavior in the deep space-like region.

\subsection{Spin 1/2 Fermions}

The  thermal Wightman functions for spin-1/2 fields are given
by
\begin{eqnarray}
S_{>\alpha\beta}(x)=&&-i{\rm
Tr}\big(\varrho\,\psi_{\alpha}(x)\overline{\psi}_{\beta}(0)\big)\nonumber\\
S_{<\alpha\beta}(x)=&&i{\rm
Tr}\big(\varrho\,\overline{\psi}_{\beta}(0)\psi_{\alpha}(x)\big)\nonumber
.\end{eqnarray}
The ordering of the spinor indices is important, but the indices
will be suppressed in the following. The relative  sign
difference between the two functions is conventional.

At zero temperature the  Wightman functions are
\begin{eqnarray}
S_{>}(x)\big|_{T=0}=&&i\gamma\cdot\partial\,{i\over 4\pi^{2}}{1\over
(t-i\epsilon)^{2} -r^{2}}\nonumber\\
S_{<}(x)\big|_{T=0}=&&i\gamma\cdot\partial\,{i\over 4\pi^{2}}
{1\over (t+i\epsilon)^{2}-r^{2}}.\nonumber
\end{eqnarray}
The former is analytic in the entire lower-half of the complex $t$ plane
and the
upper is analytic in the entire upper-half of that plane.

At nonzero temperature the regions of analyticity are reduced.
In the complex time plane, $S_{>}(x)$ is analytic in the lower strip
$-\beta\le{\rm Im}\,t\le0$ and $S_{<}(x)$ is analytic in the upper strip
$0\le{\rm Im}\,t\le\beta$. On the boundaries of these regions the values are
related by the KMS conditions \cite{KMS}:
\begin{eqnarray}
S_{>}(t-i\beta,\vec{r})=&&-S_{<}(t,\vec{r})\nonumber\\
S_{<}(t+i\beta,\vec{r})=&&-S_{>}(t,\vec{r}).\label{3cKMS}
\end{eqnarray}
The difference between the Wightman functions is the anticommutator of the
fields.
For free fields this anticommutator is a c-number:
\begin{equation}
S_{>\alpha\beta}(x)-S_{<\alpha\beta}(x)=
-i\big\{\psi_{\alpha}(x),\overline{\psi}_{\beta}(0)\big\}
\end{equation}
and thus the right hand side is independent of temperature.
This implies
\begin{displaymath}
S_{>}(x)-S_{<}(x)=S_{>}(x)\big|_{T=0}-S_{<}(x)\big|_{T=0}.
\end{displaymath}
so that the Wightman functions have the structure
\begin{eqnarray}
S_{>}(x)=&&S_{>}(x)\big|_{T=0}+E(x)\nonumber\\
S_{<}(x)=&&S_{<}(x)\big|_{T=0}+E(x).\nonumber
\end{eqnarray}
The free-field equation $-i\gamma\!\cdot\!\partial\,\psi=0$
implies that $E(x)$ must satisfy $-i\gamma\!\cdot\!\partial\,
E=0$ . To solve this, let the unknown function
$E=i\gamma\!\cdot\!\partial\,F(x)$ where $\sqr66\,F(x)=0$.
The general spherically symmetric solution is
\begin{displaymath}
E(x)=i\gamma\cdot\partial\;{1\over r}
\Big(G_{1}(u)+G_{2}(v)\Big).
\end{displaymath}
From their definitions, the Wightman functions satisfy
$[S_{>}(x)]^{\dagger}=\gamma_{0}S_{>}(-x^{*})\gamma_{0}$
and this makes the two functions the same: $G_{1}=G_{2}$.
 Combining this with the previous zero-temperature results
allows the full thermal Wightman function to be expressed as
\begin{eqnarray}
S_{>}(x)=&&(i\gamma\cdot\partial)\,\sigma_{>}(x)
\nonumber\\
S_{<}(x)=&&(i\gamma\cdot\partial)\,\sigma_{<}(x),
\label{3aS>S<}\end{eqnarray}
in which the new functions have the form
\begin{mathletters}\begin{eqnarray}
\sigma_{>}(x)=&&{-i\over 8\pi^{2}r}\bigg[{1\over u-i\epsilon}+G(u)+{1\over
v+i\epsilon}+G(v)\bigg]\\
 \sigma_{>}(x)=&&{-i\over 8\pi^{2}r}\bigg[{1\over u+i\epsilon}+G(u)+{1\over
v-i\epsilon}+G(v)\bigg],
\end{eqnarray}\label{3aBoths}\end{mathletters}
All the temperature dependence is contained in the single unknown function $G$.
It will be determined by the fermionic KMS condition Eq. (\ref{3cKMS}), which
requires that $\sigma_{>}(t-i\beta,r)=-\sigma_{<}(t,r)$. For the function
$G$ this
requires that
\begin{displaymath}
{1\over u-i(\beta+\epsilon)}+G(u-i\beta)=-\Big({1\over u+i\epsilon}+G(u)\Big).
\end{displaymath}
The solution for  $G$ is
\begin{equation}
G(u)=\sum_{n=1}^{\infty}(-1)^{n}\bigg[{1\over u-i(n\beta\!+\!\epsilon)}
+{1\over u+i(n\beta\!+\!\epsilon)}\bigg].\label{3aSeries}
\end{equation}
This is analytic in the closed strip
$-\beta\le {\rm Im}\,t\le \beta$.
The nearest poles are just above this strip at $r+t=i(\beta+\epsilon)$ and
just below the strip at $r+t=-i(\beta+\epsilon)$.
The alternating signs will produce more rapid convergence than in the bosonic
case.

\underbar{Complex time in the open strip:}
For many purposes one is interested in either in real time or
in complex time in the open strip
\begin{equation}
-\beta<{\rm Im}\,t<\beta.\label{3aStrip}
\end{equation}
For $t$ in this open region one can set $\epsilon\to\! 0$ in
 Eq. (\ref{3aSeries}) which allows  the sum to be easily
performed:
\begin{equation}
G(u)=-{1\over u}+{\pi T\over\sinh(\pi Tu)}.\end{equation}
There is, of course, no pole at $u=0$. There are poles at
$u=\pm in\beta$ because of the limt $\epsilon\to\!0$.
The  results for $\sigma_{>}$ and $\sigma_{<}$ are
\begin{mathletters}\begin{eqnarray}
\sigma_{>}(x)={1\over 8\pi r}&&\big(\delta(u)-\delta(v)\big)\nonumber\\
-{iT\over 8\pi r}&&\bigg[{1\over\sinh(\pi Tu)}+{1\over\sinh(\pi T v)}\bigg] \\
\sigma_{<}(x)={1\over 8\pi r}&&\big(-\delta(u)+\delta(v)\big)\nonumber\\
-{iT\over 8\pi r}&&\bigg[{1\over\sinh(\pi Tu)}+{1\over\sinh(\pi T v)}\bigg]
.\end{eqnarray}\label{3as>s<}\end{mathletters}
It is worth noting that the KMS condition
$\sigma_{>}(t-i\beta,r)=-\sigma_{<}(t,\beta)$ is now  satisfied in a
trivial manner
 because of the restriction to the open strip in Eq. (\ref{3aStrip}).
Because of this, it is only possible for $t$ and $t-i\beta$ to lie in the open
strip if both are complex and if both are complex then the Dirac delta
functions
have  no support.  In the full solution Eq. (\ref{3aSeries})
the KMS condition is satisfied  nontrivially.

\underbar{Imaginary time:} The imaginary time formalism uses $t=-i\tau$ where
$\tau$ is real. As occurred for bosons, the two Wightman functions
are equal, $\sigma_{>}(-i\tau,r)=\sigma_{<}(-i\tau,r)$ and given by
\begin{equation}
\sigma(-i\tau,r)={-iT\over 2\pi r}\,{\sinh(2\pi Tr)\cos(\pi T\tau)\over
\cosh(2\pi Tr)-\cos(2\pi T\tau)}.
\end{equation}
This is antiperiodic under $\tau\to\tau\pm\beta$.

\underbar{Arbitrary complex time:} Without approximation one can sum the series
in Eq. (\ref{3aSeries}) to obtain
\begin{eqnarray}
G(u)=&&i{T\over 2}\psi[{1\over 2}\!+\!{T\over 2}(\epsilon\!+\!iu)]
-i{T\over 2}\psi[1+{T\over 2}(\epsilon\!+\!iu)]\nonumber\\
-&&i{T\over 2}\psi[{1\over 2}+{T\over 2}(\epsilon\!-\!iu)]
+i{T\over 2}\psi[1\!+\!{T\over 2}(\epsilon\!-\!iu)].\nonumber
\end{eqnarray}
Since $\psi(z)=d\ln[\Gamma(z)]/dz$ this is equivalent to
\begin{displaymath}
G(u)={\partial\over\partial r}\ln[g(u)].\end{displaymath}
in which the lower-case function $g$ is
\begin{mathletters}\begin{equation}
g(u)={\Gamma[{1\over 2}\!+\!{T\over 2}(\epsilon+iu)]
\Gamma[{1\over 2}\!+\!{T\over 2}(\epsilon-iu)]\over
\Gamma[1\!+\!{T\over 2}(\epsilon+iu)]
\Gamma[1\!+\!{T\over 2}(\epsilon-iu)]}
\end{equation}
Putting this together in Eq. (\ref{3aBoths}) gives the general result
\begin{eqnarray}
\sigma_{>}(x)=&&\sqr66\,s_{>}(x)\nonumber\\
\sigma_{<}(x)=&&\sqr66\,s_{<}(x).
\end{eqnarray}
in which the lower-case functions are
\begin{eqnarray}
s_{>}(x)=&&-{i\over 16\pi^{2}}\ln\bigg[{g(u)g(v)\over (u-i\epsilon)
(v+i\epsilon)}\bigg]\nonumber\\
s_{<}(x)=&&-{i\over 16\pi^{2}}\ln\bigg[{g(u)g(v)\over (u+i\epsilon)
(v-i\epsilon)}\bigg].
\end{eqnarray}\label{3aSigma}\end{mathletters}
The next section will show will express the Wightman functions for
gravitinos in terms of these functions.

\subsection{Spin 3/2 Gravitinos}

The massless Rarita-Schwinger field \cite{JS} plays an
important role in supergravity as the spin 3/2,
supersymmetric partner of the graviton. In that context it is
referred to as the gravitino. As occurred in the bosonic
cases, the passage from the lower spin fermion (1/2) to the
higher spin fermion (3/2) is easily accomplished.  The free
Lagrangian density for the field $\psi_{\mu}(x)$ is \cite{BSD}
\begin{displaymath}
{\cal L}=-\epsilon^{\mu\nu\alpha\beta}\;\overline{\psi}_{\mu}
\gamma_{\nu}\gamma_{5}\partial_{\alpha}\psi_{\beta}
+{\cal L}_{g.f.}.
\end{displaymath}
The gauge fixing is necessary so as to break the invariance of
the first term under transformations $\psi_{\beta}
\to \psi_{\beta}+\partial_{\beta}\psi$, where $\psi$ is any
spin-1/2 field. A convenient choice for gauge-fixing is
\begin{displaymath}
{\cal L}_{g.f.}={i\over
2}\,\overline{\psi}^{\nu}\!\gamma_{\nu}
(\gamma\cdot\partial)\gamma_{\lambda}\psi^{\lambda}
\end{displaymath}
The free Lagrangian can be rewritten  as
\begin{equation}
{\cal L}=-\overline{\psi}^{\mu}L_{\mu\nu}\psi^{\nu}
\label{3bLag},
\end{equation}
where the tensor $L_{\mu\nu}$ is linear in the first
derivatives:
\begin{equation}
L^{\mu\nu}=-i(\gamma^{\mu}\partial^{\nu}
\!+\!\gamma^{\nu}
\partial^{\mu}\!+\!\eta^{\mu\nu}\gamma\!\cdot\!\partial
\!-\!{1\over
2}\gamma^{\mu}(\gamma\!\cdot\!\partial)\gamma^{\nu}).
\label{3bL}\end{equation}
Each Wightman functions is a 4$\times$ 4 matrix in spinor
space and a rank 2 tensor in the Lorentz indices:
\begin{eqnarray}
S_{>}^{\;\mu\nu}(x)=&&-i{\rm
Tr}\big(\varrho\,\psi^{\mu}(x)
\overline{\psi}^{\,\nu}(0)\big)\nonumber\\
S_{<}^{\;\mu\nu}(x)=&&
i{\rm Tr}\big(\varrho\,
\overline{\psi}^{\,\nu}(0)\psi^{\mu}(x)\big).\label{3bS>S<}
\end{eqnarray}
The spinor indices
 have been  suppressed.
The relative sign difference in the definitions coincides with the
spin 1/2 convention.
As in the bosonic case  the most efficient way to proceed is
to display the answer immediately and then perform checks.
In this gauge the Wightman functions  can be written in terms
of the same matrix-differential operator
$L^{\mu\nu}$
\begin{eqnarray}
S^{\mu\nu}_{>}(x)=&&L^{\mu\nu}\sigma_{>}(x)
\nonumber\\
S^{\mu\nu}_{<}(x)=&&L^{\mu\nu}\sigma_{<}(x),\label{3bS>S<}\end{eqnarray}
where $\sigma_{>}(x)$ and $\sigma_{<}(x)$ are the basic spin
1/2 functions given in Eq. (\ref{3aSigma}). The subsequent discussion
will confirm these  results.

(1) The spin 1/2 functions $\sigma_{>}(x)$ and
$\sigma_{<}(x)$ guarantee that the gravitino Wightman
functions in Eq. (\ref{3bS>S<})  satisfy the KMS conditions
\begin{eqnarray}
S_{>}^{\mu\nu}(t-i\beta,\vec{r})&&=-S_{<}^{\mu\nu}(t,\vec{r})\nonumber\\
S_{<}^{\mu\nu}(t+i\beta,\vec{r})&&=-S_{>}^{\mu\nu}(t,\vec{r}),
\end{eqnarray}
and have the correct analyticity properties.

(2) The equation of motion for the gravitino which follows
directly from Eq. (\ref{3bLag})  is
\begin{displaymath}
L_{\alpha\mu}\psi^{\mu}(x)=0.
\end{displaymath}
Therefore the Wightman functions should satisfy
\begin{displaymath}
L_{\alpha\mu}S^{\mu\nu}_{>}(x)
=L_{\alpha\mu}S^{\mu\nu}_{<}(x)=0.
\end{displaymath}
This is easily satisfied because of the identity
\begin{equation}
L_{\alpha\mu}L^{\mu\nu}=-\delta_{\alpha}^{\;\mu}\,\sqr66
\end{equation}
and the fact that $\sqr66\,\sigma_{>}=0$ and
$\sqr66\,\sigma_{<}=0$.

(3) The third check of Eq. (\ref{3bS>S<}) is that the
gravitino field operators must satisfy the canonical
commutation relations at equal time. For this it is convenient
to express the partial derivative of the Lagrangian as
\begin{displaymath}
{\partial{\cal L}\over
\partial(\partial^{\rho}\psi^{\lambda})}
=\overline{\psi}^{\nu}\Gamma_{\nu\lambda\rho},
\end{displaymath}
where $\Gamma_{\nu\lambda\rho}$ is the matrix
\begin{equation}
\Gamma_{\nu\lambda\rho}\equiv
i(\gamma_{\nu}\eta_{\lambda\rho}+\gamma_{\lambda}
\eta_{\nu\rho}
-\eta_{\nu\lambda}\gamma_{\rho}-
{1\over 2}\gamma_{\nu}\gamma_{\rho}\gamma_{\lambda}).
\end{equation}
Note that
$L_{\nu\lambda}=-\Gamma_{\nu\lambda\rho}\partial^{\rho}$. The
canonical momentum conjugate to $\psi^{\lambda}(x)$ is
\begin{displaymath}
\Pi_{\lambda}(x)=\overline{\psi}^{\nu}\Gamma_{\nu\lambda 0}.
\end{displaymath} These momenta obey the canonical
equal-time anticommutation relations
\begin{displaymath}
-i\Big\{\psi^{\mu}(x),\Pi_{\lambda}(0)\Big\}_{t=0}
=-\delta_{\lambda}^{\;\mu}\,\delta^{3}(\vec{r}),
\end{displaymath}
so that the Wightman functions must satisfy
\begin{equation}
\Big[(S_{>}^{\mu\nu}(x)-S_{<}^{\mu\nu}(x))
\Gamma_{\nu\lambda
0}\Big]_{t=0}=-\delta_{\lambda}^{\;\mu}\,\delta^{3}(\vec{r}).
\label{3bMom}\end{equation}
To check that these are satisfied by Eq. (\ref{3bS>S<}), first
multiply by the matrix $\Gamma$ to obtain
\begin{eqnarray}
(S_{>}^{\mu\nu}(x)-S_{<}^{\mu\nu}(x))
\Gamma_{\nu\lambda\rho}
=\big(&&\delta_{\lambda}^{\;\mu}\delta_{\rho}^{\;\kappa}
-\delta_{\rho}^{\;\mu}\delta_{\lambda}^{\;\kappa}
+\eta_{\lambda\rho}\,\eta^{\mu\kappa}\nonumber\\
-i&&\gamma_{5}\epsilon_{\lambda\rho}^{\;\;\;\;\mu\kappa}
\big)\partial_{\kappa}(\sigma_{>}(x)-\sigma_{<}(x)).\nonumber
\end{eqnarray}
For canonical momenta, set $\rho=0$ and the time $t=0$.
Because $\sigma_{>}(0,\vec{r})=\sigma_{<}(0,\vec{r})$ only
the time derivatives on the right (i.e. $\kappa=0$) are
nonvanishing. This reduces to
\begin{displaymath}
\Big[(S_{>}^{\mu\nu}(x)-S_{<}^{\mu\nu}(x))
\Gamma_{\nu\lambda
0}\Big]_{t=0}=\delta_{\lambda}^{\;\mu}
\big[\dot{\sigma}_{>}(x)-\dot{\sigma}_{<}(x)\big]_{t=0}.
\end{displaymath}
Thus Eq. (\ref{3bMom}) is satisfied because of the properties
of the functions $\sigma_{>}(x)$ and $\sigma_{<}(x)$ from Sec IIIA.

\underbar{General Covariant Gauge:} The results for the
gravitino have been displayed in a particular gauge.
As was done for the graviton in Sec IIC, one can add a more
general gauge-fixing term than used above.  The Wightman
functions can still be expressed in terms of two derivatives
of the functions $\sigma_{>}(x)$ and $\sigma_{<}(x)$ in Eq.
(\ref{3as>s<}).

\section{Discussion}

It is rather surprising that the large distance effects of massless
particles are, in some sense, simpler at $T>0$ than in vacuum.
At zero temperature the  Wightman function for spinless bosons is
\begin{equation}
D_{>}(x)\big|_{T=0}={i\over 4\pi^{2}}{1\over
(t-i\epsilon)^{2}-r^{2}},\end{equation}
and  for spin 1/2 fermions
is $S_{>}(x)=i\gamma^{\mu}\partial_{\mu}\sigma_{>}(x)$, where
\begin{equation}
\sigma_{>}(x)\big|_{T=0}={i\over 4\pi^{2}}{1\over
(t-i\epsilon)^{2}-r^{2}}.\end{equation}
In perturbative calculations, the slow fall-off at large distances
produces long range correlations both in the deep space-like and the deep
time-like directions. At $T>0$ the situation is very different
as  indicated below.

\underbar{Deep space-like region:} For $r-|t|\gg 1/(\pi T)$ the asymptotic
behavior of Eqs. (\ref{2aD>D<})  is
\begin{equation}
D_{>}(x)\to{-iT\over 4\pi r} - {iT\over 2\pi r}e^{-2\pi T r}
\cosh(2\pi T t)+\dots\end{equation}
This behavior can also be understood from the Matsubara formalism,
\cite{b1,b2,b3} which has discrete frequencies $\omega_{n}=2n\pi T$.
The static, $n=0$ mode is responsible for the leading $T/r$ behavior. Each
higher mode produces an $\exp(-\omega_{n}r)$ fall-off.
For fermions the asymptotic behavior of Eq. (\ref{3as>s<}) is
\begin{equation}
\sigma_{>}(x)\to {-iT\over 2\pi r}e^{-\pi Tr}\cosh(\pi Tt)+\dots
\end{equation}
Since fermion have odd frequencies, $\omega_{n}=(2n+1)\pi T$, the
exponential fall-off also comes from $\exp(-\omega_{n}r)$.

\underbar{Deep time-like region:} In the deep time-like region defined by
$|t|-r\gg 1/(\pi T)$, the asymptotic behaviors of Eqns. (\ref{2aD>D<})
and (\ref{3as>s<}) are
\begin{eqnarray}
D_{>}(x)\to&& {-iT\over 2\pi r}e^{-\pi T|t|}\sinh(2\pi Tr)+\dots
\nonumber\\\nonumber\\
\sigma_{>}(x)\to&&{iT\over 2\pi r}e^{-\pi T|t|}\sinh(2\pi Tr)+\dots
\end{eqnarray}
It is not obvious why
both Wightman functions should fall exponentially in the
time-like region and with the same exponent.

All the Wightman functions for higher spin are
simply related (by derivatives) to these basic functions.
It thus appears that in comparison with zero temperature, the quantitatively
large effects of  massless particles at non-zero temperature comes
predominantly from the $T/r$ behavior of bosons at large spatial separation
and from  the light cone $t^{2}=r^{2}$.  Subsequent publications will
explore the physical consequences of these results.

\acknowledgments
This work was supported in part by the U.S.  National Science Foundation under
grant PHY-9900609.

\end{document}